\documentclass[aps,prl,twocolumn]{revtex4-1}
\usepackage{graphicx}
\usepackage{hyperref}
\usepackage{amsmath}
\usepackage{amssymb}

\usepackage{cancel}

\newcommand{\beq}{\begin{eqnarray}}
\newcommand{\eeq}{\end{eqnarray}}
\newcommand{\bea}{\begin{eqnarray}}
\newcommand{\eea}{\end{eqnarray}}
\newcommand{\mpl}{M_{\rm pl}}

\newcommand{\Tf}{T_{\rm f}}
\newcommand{\Td}{T_{\rm D}}
\newcommand{\Tw}{T_{\rm W}}

\begin{document}
\begin{flushright}
UMD-PP-012-024\\
\end{flushright}
\title{Baryogenesis for WIMPs}

\author{Yanou Cui}
\affiliation{Maryland Center for Fundamental Physics, Department of Physics, University of Maryland, College Park, MD 20742, USA}
\author{Raman Sundrum}
\affiliation{Maryland Center for Fundamental Physics, Department of Physics, University of Maryland, College Park, MD 20742, USA}
\preprint{UMD-PP-xxxxxx}

\begin{abstract}
    We propose a robust, unified framework, in which the similar baryon and dark matter cosmic abundances both arise from the physics of weakly interacting massive particles (WIMPs), with the rough quantitative success of the so-called ``WIMP miracle''.
    In particular the baryon asymmetry arises from the decay of a meta-stable WIMP after its thermal freezeout at or below the weak scale. A minimal model and its embedding in R-parity violating (RPV) SUSY are studied as examples. The new mechanism saves RPV SUSY from the potential crisis of washing out primordial baryon asymmetry. Phenomenological implications for the LHC and precision tests are discussed.

\end{abstract}


\maketitle

\section{Introduction}
The observed dark matter (DM) and baryon abundances $\Omega_{\rm DM}\simeq23\%, \Omega_B\simeq4\%$ have long been addressed with separate mechanisms at separate scales. The conventional paradigm for DM theory is the ``WIMP miracle'' which gives a striking yet rough guideline for $\Omega_{\rm DM}$: thermal relic abundance of a stable WIMP naturally falls in the right ballpark of the observed $\Omega_{\rm DM}$. The past few years have seen rising interest in the intriguing ``coincidence'' of $\Omega_{\rm DM}\sim \Omega_B$, bringing in the new paradigm of ``Asymmetric Dark Matter'' (ADM)\cite{ADM}. However, ADM's success is at the cost of the WIMP miracle. A unified mechanism that can both address the ``coincidence'' and preserve the WIMP miracle would surely be more desirable. Only very recently, a few attempts have been made in this direction\cite{McDonald:2011zz,{Cui:2011ab},Davidson:2012fn}. However, \cite{McDonald:2011zz} is highly sensitive to multiple initial conditions; \cite{Cui:2011ab} proposes a novel baryogenesis triggered by WIMP DM annihilation, but moderate adjustment of parameters is required to suppress washout effects; \cite{Davidson:2012fn} is also sensitive to washout, and its reliance on leptogenesis further restricts working parameters. In this paper we explore an alternative baryogenesis mechanism with a robust connection to the WIMP miracle and less sensitivity to model details. \\
\indent Various scenarios addressing the electroweak Hierarchy Problem come with new particles of WIMP type. Generically there may be an  array of WIMPs, some of which are stable, some of which decay promptly, some of which have long lifetimes, depending on protection from symmetries and mass hierarchies. Although conventionally, the WIMP miracle only applies to stable WIMPs as DM candidates, it has more general application. We consider a \textit{meta-stable WIMP} that first undergoes thermal freezeout and later decays in a $\cancel{B}, \cancel{CP}$ way, triggering baryogenesis\cite{Sakharov:1967dj}. The resultant $\Omega_B$ inherits the would-be miracle abundance from the WIMP parent up to only moderate suppression from CP asymmetry and baryon/WIMP mass ratio, and thus makes it roughly comparable to $\Omega_{\rm DM}$ of a WIMP DM. A precise fit to $\Omega_B, \Omega_{\rm DM}$ only requires $O(1)$ adjustment of the different WIMP parameters, and is insensitive to the precise WIMP lifetime. Furthermore when embeded in RPV $\cancel{B}$ SUSY, this mechanism provides a remedy to a cosmological problem there: $\cancel{B}$ leading to prompt decays at collider typically washes out primordial baryon density and calls for baryogenesis below the weak scale. Alternative solutions to this problem \cite{Dreiner:1992vm, Davidson:1996hs} are less generic. Refs.\cite{Dimopoulos:1987rk,{Cline:1990bw}} considered low scale baryogenesis in $\cancel{B}$ SUSY to solve the gravitino problem, but the results are sensitive to details about the inflaton or gravitino. These works do not address the WIMP miracle or $\Omega_{\rm DM}-\Omega_B$ ``coincidence''.
\section{General formulation}
{\underline{Stage-1: WIMP freezeout}\\
 A thermal WIMP $\chi$ freezes out of equilibrium around $T_{\rm f}$ when its thermal annihilation rate $\Gamma_A\simeq n_\chi^{\rm eq}\langle\sigma_{\rm A} v\rangle$ matches Hubble rate $H$. This results in the estimate\cite{Kolb:1990vq}: 
 \beq
T_f\simeq m_\chi\left[\ln\left(0.038({g}/{g_*^{1/2}})m_\chi\mpl\langle\sigma_{\rm A} v\rangle\right)\right]^{-1},\label{wimpfo}
\eeq
which is typically $\sim\frac{1}{20}m_\chi$. $g$ counts the internal degrees of freedom of $\chi$. $g_*$ counts total degrees of freedom of relativistic species. At the end of this stage the co-moving density of $\chi$ is:
\bea
    Y_\chi(\Tf)&=&\frac{n_\chi^{\rm eq}(\Tf)}{s(\Tf)}\simeq 3.8\frac{g_*^{1/2}}{g_{*s}}\frac{m_\chi}{\Tf}\left(m_\chi\mpl\langle\sigma_{\rm A} v\rangle\right)^{-1}, \label{chiinitial}
\eea
where $s$ is entropy, $g_{*s}\equiv\frac{45}{2\pi^2}\frac{s}{T^3}$.
If $\chi$ is stable, $Y_\chi(\Tf)\simeq Y_\chi(T_0)$, where $T_0$ is today's temperature, and its relic density today is:
\bea
   \nonumber
   \Omega_{\chi}&=&\frac{m_\chi Y_\chi(\Tf)s_0}{\rho_0}\simeq0.1\frac{\alpha_{\rm weak}^2/(\rm TeV)^2}{\langle\sigma_{\rm A} v\rangle}\\
   &\simeq&0.1\left(\frac{g_{\rm weak}}{g_\chi}\right)^4\left(\frac{m_{\rm med}^4}{m_\chi^2\cdot\rm TeV^2 }\right),
   \label{omegawimp}
\eea
where $\rho_0=\frac{3H_0^2}{8\pi G}$, $H_0$ and $s_0$ are the current energy density, Hubble rate and entropy, respectively.
The second line in eq.(\ref{omegawimp}) manifests the dependence on model parameters in the generic case of heavier mediator with $m_{\rm med}\gtrsim m_\chi$. Now consider two species of WIMPs: $\chi_{\rm DM}$ which is stable DM, and $\chi_{B}$ which decays at time $\tau$, after freezeout. The observation that eq.(\ref{omegawimp}) readily fits the measured dark matter abundance $\Omega_{\chi_{\rm DM}}\simeq23\%$ is the well-known \textit{``WIMP miracle''}.
In case of $\chi_B$, $Y_{\chi_B}(\Tf)\equiv Y_{\chi_B}^{\rm ini}$, acts as the initial condition for later baryogenesis, as we now discuss.\\
{\underline{Stage-2: Baryogenesis}}\\
 Consider the baryogenesis ``parent'' $\chi_B$ to have \cancel{CP}, \cancel{B} decay after its freezeout but before BBN, i.e. $1~\rm MeV\sim T_{\rm BBN}<\Td<\Tf$, so that we can treat freezeout and baryogenesis as nearly decoupled processes, and retain the conventional success of BBN.
Solving the Boltzmann equations\cite{Kolb:1990vq} we get the asymmetric baryon density per co-moving volume today $Y_{{B}}(T_0\approx 0)$:
\bea\nonumber
Y_{{B}}(0)&=&\epsilon_{\rm CP}\int_0^{T_{\rm D}}\frac{dY_{\chi_B}}{dT}\exp\left(-\int^{T}_0 \frac{\Gamma_{\rm W}(T')}{H(T')}\frac{dT'}{T'}\right)dT\\ &+& Y_B^{\rm ini}\exp\left(-\int^{T_{\rm \rm ini}}_0 \frac{\Gamma_{\rm W}(T)}{H(T)}\frac{dT}{T}\right), \label{Bsol}
\eea
where we assume $\chi_B$ decay violates $B$ by 1 unit. $\epsilon_{\rm CP}$ is CP asymmetry in $\chi_B$ decay, $\Gamma_{\rm W}$ is the rate of $\cancel{B}$ washout processes. $Y_B^{\rm ini}$ represents possible pre-existing B-asymmetry, which we first assume to be $0$.
 In case of weak washout, i.e., $\Gamma_{\rm W}<H$, the exponential factor in eq.(\ref{Bsol}) can be dropped. Then using eqs.(\ref{omegawimp},\ref{Bsol}) we obtain:
\beq
  Y_{{B}}(0)\simeq\epsilon_{\rm CP} Y_{\chi_B}(\Tf), ~~ \Omega_B(0)=\epsilon_{\rm CP}\frac{m_p}{m_{\chi_B}}\Omega_{\chi_B}^{\tau\rightarrow\infty}\label{omegab},
\eeq
where $\Omega_{\chi_B}^{\tau\rightarrow\infty}$ is the would-be relic abundance of WIMP $\chi_B$ in the limit it is stable, given by eq.(\ref{omegawimp}). $\Omega_B$ given in eq.(\ref{omegab}) is insensitive to the precise lifetime of $\chi_B$ as long as it survives thermal freezeout. The observed $\Omega_B\simeq4\%$ today corresponds to $Y_{{B}}(0)\equiv \frac{n_B}{s}\simeq10^{-10}$.
 $\Omega_B(0)$ in eq.(\ref{omegab}) takes the form of WIMP miracle, but with an extra factor $\epsilon_{\rm CP}\frac{m_p}{m_{\chi_B}}\sim10^{4}-10^{-3}$ for weak scale $\chi_B$ and $O(1)$ couplings and phases. Nonetheless as can be seen from eq.(\ref{omegawimp}), the observed $\frac{\Omega_B}{\Omega_{\rm DM}}\approx\frac{1}{5}$ can readily arise from $O(1)$ difference in masses and couplings associated with the two WIMP species $\chi_{\rm DM}$ and $\chi_B$. This is our central result. \\
\indent  Note that as long as $\chi$ decays well before BBN, the produced baryons get thermalized efficiently, because $\Gamma_{pX\rightarrow pX}\sim T\gg H$ at $T_{\rm BBN}\ll T\lesssim T_{\rm EW}$, where $X$ can be $e^\pm, p, \bar{p}$ in the thermal bath. Thus as in conventional baryogenesis, the symmetric component of baryons is rapidly depleted by thermal annihilation. Dilution/reheating from $\chi_B$ decay is negligible because at $\Td$ the energy density of $\chi_B$ is much less than radiation density. To see this, recall that today $T_0\approx10^{-4}\rm eV$, $\frac{\Omega_B(T_0)}{\Omega_{\rm rad}(T_0)}\approx10^{3}$. Red-shifting back to $\Td$ and using eq.(\ref{omegab}) we get $\frac{\Omega_{\chi_B}(\Td)}{\Omega_{\rm rad}(\Td)}\approx\frac{\Omega_B(T_0)}{\Omega_{\rm rad}(T_0)}\frac{m_{\chi_B}}{\epsilon_{\rm CP}m_p}\frac{T_0}{\Td}\ll1$ for $\Td>T_{\rm BBN}$ and sizeable $\epsilon_{\rm CP}$.
\section{Minimal model and constraints}
 We add to the Standard Model (SM) Lagrangian:
 \bea
\Delta\mathcal{L}&=&\nonumber\lambda_{ij}\phi d_id_j + \varepsilon_i \chi\bar{u}_i\phi +M_{\chi}^2\chi^2 + y_i\psi\bar{u}_i\phi +M_{\psi}^2\psi^2\\ &+& \alpha\chi^2S + \beta|H|^2 S + M_S^2S^2 + \rm h.c.\label{minimalmodel}
\eea
where  $H$ is the SM Higgs, $d, u$ are RH SM quarks, with family indices $j=1,2,3$, $\phi$ is a di-quark scalar with same SM gauge charge as $u$. 
$\chi, \psi$ are SM singlet Majorana fermions, and $S$ is a singlet scalar. $\chi\equiv\chi_B$ is the earlier WIMP parent for baryogenesis. $\varepsilon_i\ll1$ are our formal small parameters leading to long-lived $\chi$. They can represent a naturally small breaking of a $\chi$-parity symmetry under which only $\chi$ is odd.  $S$ mediates thermal annihilation of $\chi\chi$ into SM states.
 The first 3 terms of eq.(\ref{minimalmodel}) give rise to collective breaking of $U(1)_B$. Out-of-equilibrium decay $\chi\rightarrow\phi^* {u}$ is followed by the prompt decay $\phi\rightarrow{d}{d}$ with $\Delta B=1, \epsilon_{\rm CP}\neq0$.
 CP asymmetry $\epsilon_{\rm CP}$ in $\chi$ decay comes from the $\psi$-mediated interference between tree-level and loop diagrams as shown in Fig.(\ref{fig:decaycpv}). In the case of $M_{\psi}>M_{\chi}$, in close analogy to leptogenesis\cite{Chen:2007fv}, we obtain:
\beq
 \epsilon_{\rm CP}\simeq\frac{1}{8\pi}\frac{1}{\sum_i|\varepsilon_i|^2}Im\left\{\left(\sum_i\varepsilon_iy_i^{*}\right)^2\right\}\frac{M_{\chi}}{M_{\psi}}\label{cpv},
\eeq
which is non-zero for generic complex couplings. We also see that the key to a large $\epsilon_{\rm CP}$ is to have $y_i\sim O(1)$ for at least one flavor $i$. Note the analogous $\epsilon_{\rm CP}$ from $\psi$ decay is $O(\varepsilon^2)$, with $\varepsilon\leftrightarrow y, M_\chi\leftrightarrow M_\psi$ in eq.(\ref{cpv}).\\
\indent It is straightforward to incorporate WIMP DM by introducing another singlet $\chi_{\rm DM}$ with analogous interactions to $\chi$, except with $\varepsilon_{\rm DM}=0$ enforced by an exact $\chi_{\rm DM}$-parity. We will not write out the $\chi_{\rm DM}$ physics explicitly. We next consider various constraints on this minimal model. We start with a generic flavor structure, and drop family indices in $y, \varepsilon$ for now. \\
\underline{Lifetime of $\chi$:} \\
 The decay rate of $\chi$ at $T<m_\chi$ is $\Gamma_{\rm D}\simeq\frac{\varepsilon^2m_{\chi}}{8\pi}$.
With $T_{\rm f}\sim100\rm~ GeV$, our requirement of $\chi$ decay within range $T_{\rm BBN}< \Td< \Tf$ leads to the constraint:
$ 10^{-13}\lesssim \varepsilon\lesssim10^{-8}$. \\
\\ 
\begin{figure}[t]
\begin{center}
\includegraphics[height=2.2 cm]{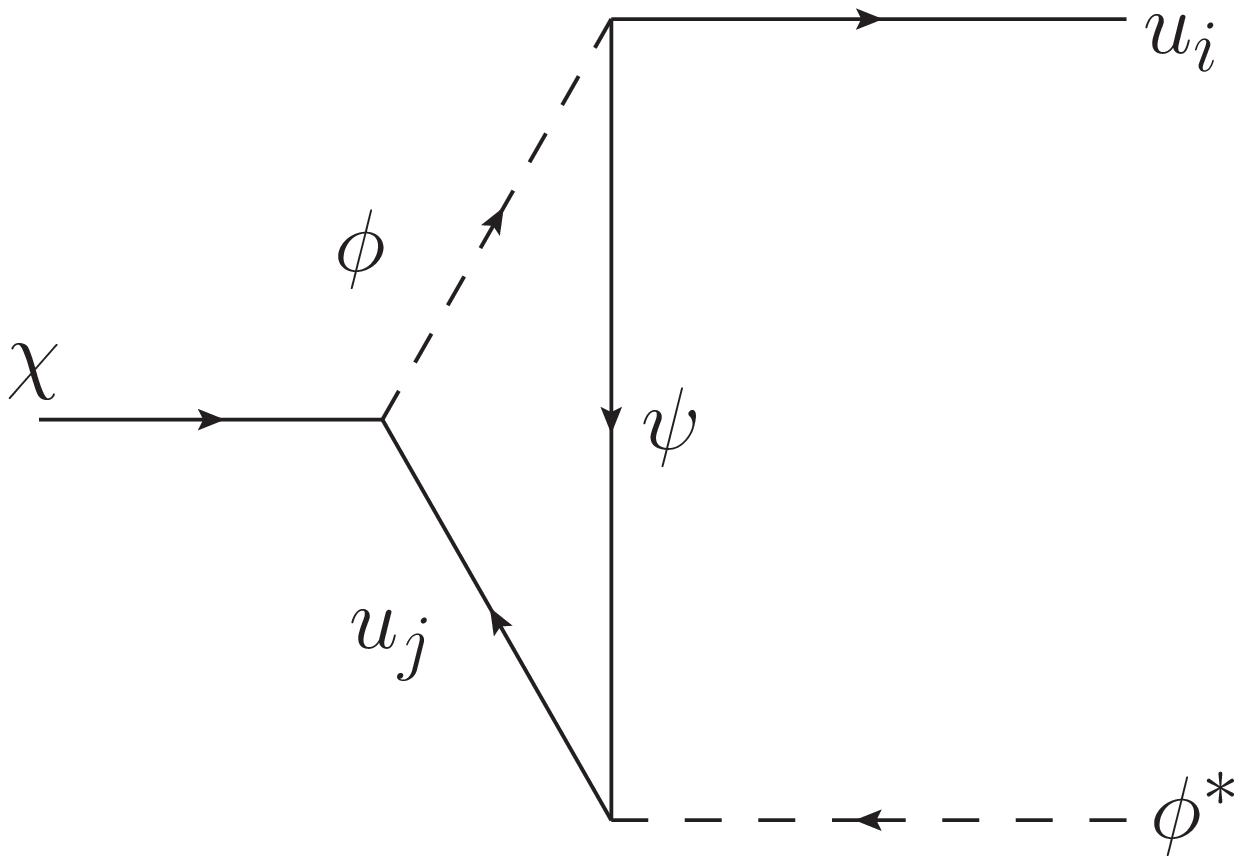}
\includegraphics[height=2.2 cm]{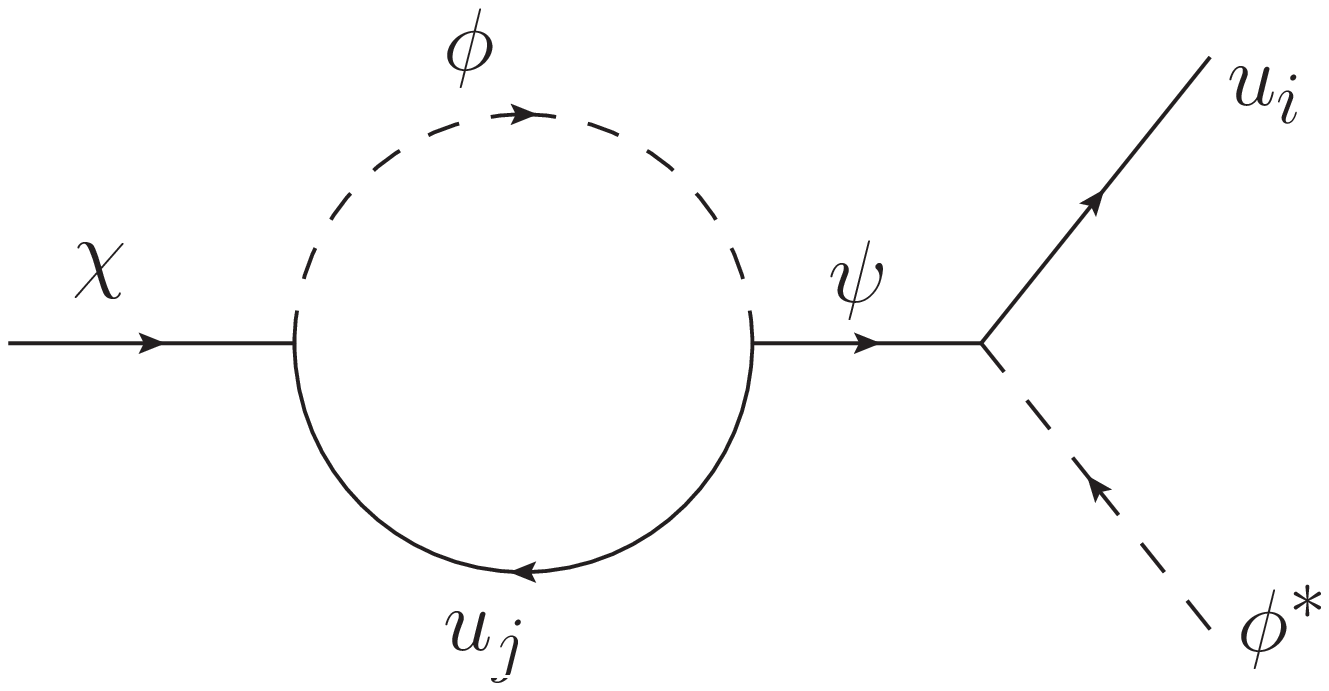}
\caption{Loop diagrams that interfere with tree-level decay to generate $\epsilon_{CP}$}
\label{fig:decaycpv}
\end{center}
\end{figure}
We next consider potential washout effects. We will focus on considering processes involving $\psi$; there are analogous diagrams with $\psi\rightarrow\chi$, but they give much looser constraints since $\varepsilon\ll y\sim O(1)$.
\underline{Early time washout:  at $T> \Lambda_{QCD}$}\\
As we will see, in this epoch $\Gamma_W/H$ decreases with $T$. Thus for each early washout process $X$, we define $\Tw^X$ such that $\Gamma_W^X\lesssim H$ for $T<\Tw^X$. We require $T_D<T_W^X$ to have weak washout effect.\\
\textbf{A.} Inverse decay $ u d d\rightarrow \psi$ via an onshell ${\phi}^*$:
\beq
   \Gamma_W^{\rm ID,\psi}\simeq \frac{n_{\psi}^{\rm eq}}{T^3}\Gamma_{\rm D, \psi}\simeq\frac{n_{\psi}^{\rm eq}}{T^3}\frac{y^2m_{\psi}}{8\pi}.
\eeq
 This gives the constraint:
\beq
   T_{\rm D}<\Tw^{ID, \psi}\simeq m_\psi\left[\ln\left(\frac{0.076}{g_*^{1/2}}\frac{y^2\mpl}{8\pi m_\psi}\right)\right]^{-1}.\label{Tinvdec}
\eeq

\textbf{B.} $\Delta{B}=1$, $2\rightarrow2$ scattering $\psi u\rightarrow\bar{d}\bar{d}$ via $\phi$-exchange:

\beq
  \Gamma_{W}^{\Delta{B}=1}\simeq \frac{y^2\lambda^2}{16\pi m_{\psi}^2}n_{\psi}^{\rm eq}, ~~{\rm for}~~ m_\psi>m_\phi,
\eeq

\beq
 T_{\rm D}<T_{\rm W}^{\Delta{B}=1}\simeq m_{\psi}\left[\ln\left(\frac{0.076}{g_*^{1/2}}\frac{\lambda^2y^2\mpl}{16\pi m_{\psi}}\right)\right]^{-1}.\label{dB1wash}
\eeq
\textbf{C.} $\Delta{B}=2$ $3\rightarrow3$  scattering $udd\rightarrow\bar{u}\bar{d}\bar{d}$ via on-shell $\phi$ and $\psi$-exchange. This is effectively $2\rightarrow2$ ($\phi^* u\rightarrow\phi\bar{u}$), and similarly to case B:
\beq
 T_{\rm D}<T_{\rm W}^{\Delta{B}=2, 2\rightarrow2}\simeq m_{\phi}\left[\ln\left(\frac{0.076}{g_*^{1/2}}\frac{y^4\mpl}{16\pi m_{\psi}}\right)\right]^{-1}.\label{22dB2wash}
\eeq

\textbf{D.} $\Delta{B}=2$ $3\rightarrow3, 2\rightarrow4, 4\rightarrow2$ scattering: $udd\rightarrow\bar{u}\bar{d}\bar{d}$ via $\psi$-exchange and offshell $\phi$, or $ud\rightarrow\bar{u}\bar{d}\bar{d}\bar{d}$.

Taking $3\rightarrow3$ for example:
\beq
  \Gamma_W^{3\rightarrow3}\sim \frac{\lambda^4y^4}{16\pi(2\pi)^3}\frac{T^{10}}{m_\phi^8m_{\psi}^2}T,
\eeq

\beq
T_{\rm D}<T_{\rm W}^{3\rightarrow3}\simeq\left(\frac{1.66g_*^{\frac{1}{2}}128\pi^4m}{y^4\lambda^4\mpl}\right)^{\frac{1}{9}}m\sim\frac{m}{20(y\lambda)^{\frac{4}{9}}}\label{33dB2wash},
\eeq
where we simplified the expression by taking all masses $\sim m$.\\
 \indent We compare the constraints on $\Td$ given in eqs. (\ref{Tinvdec}, \ref{dB1wash}, \ref{22dB2wash}, \ref{33dB2wash}) with $\Tf$ given in eq.(\ref{wimpfo}) where for this model p-wave annihilation $\langle\sigma_{\rm A} v\rangle\sim\frac{ m_\chi^2}{16\pi m_S^4}v^2$ for $m_\chi<m_S$ and $O(1)$ couplings, $v^2\sim \frac{T_f}{m_\chi}$.  With non-hierarchical weak scale masses of $\chi, \psi, \phi, S$ and $O(1)$ couplings, we find for all washout processes considered, $\Tw\sim \Tf$. Therefore with $\Td<\Tf$, early washout is not a concern. Notice that potential washout from EW sphaleron is also easily avoided since sphaleron shuts off at $\sim100~\rm GeV\gtrsim\Tf>\Td$ for $m_\chi$ up to $O(1)$TeV.

After the QCD phase transition, the neutron and proton become new effective degrees of freedom to consider. $n-\bar{n}$ oscillation is the typical washout process in this era.  The general formula for the transition probability is \cite{Mohapatra:2009wp}:
\bea
P_{n\rightarrow\bar{n}}(t)=\frac{4\delta m^2}{\Delta E^2+4\delta m^2}\sin^2(\frac{\sqrt{\Delta E^2+4\delta m^2}}{2}\cdot t)\label{Pnosci-gen}
\eea
where $\delta m$ is the \cancel{B} Majorana mass. The splitting $\Delta E\equiv E_n-E_{\bar{n}}$ is $0$ in vacuum or in medium where $n,\bar{n}$ are symmetric, e.g. thermal bath shortly after QCD transition when baryons are dominated by the symmetric component. $\Delta E\gg\delta m$ may occur in an asymmetric medium, e.g. the thermal bath close to BBN time or the nucleus environment after BBN, which strongly suppresses $P_{n\rightarrow\bar{n}}$. In a medium where there is a characteristic time scale $\tau$, the washout rate can be estimated as
\beq
  \Gamma^{n\rightarrow\bar{n}}_W\simeq P_{n\rightarrow\bar{n}}(\tau)/\tau. \label{Pnosci-med}
\eeq
\underline{Intermediate-time washout: $T\lesssim\Lambda_{QCD}$}\\
In this epoch $n$ scatters off the thermal background and $\tau$ is set by the mean free path of $n$, bound by $H^{-1}$ from above. In reality both $\Delta E$ and $\tau$ are varying functions in this period. To simplify we consider the most ``dangerous'' limit where $\Delta E\rightarrow0$ and $\tau\rightarrow H^{-1}$ which maximizes washout according to eqs.(\ref{Pnosci-gen},\ref{Pnosci-med}), $\Gamma^{n\rightarrow\bar{n}, \rm intm}_W\simeq(\delta m)^2H^{-1}$. Requiring $\Gamma^{n\rightarrow\bar{n}, \rm intm}_W<H$ at $T\lesssim\Lambda_{QCD}$, we find $\delta m\lesssim10^{-25}\rm GeV$. \\
\underline{Late-time Washout: $T<T_{BBN}$}\\
After BBN, $n$ is bound in the nucleus. Now the characteristic time $\tau$ is set by nuclear time scale which is $\tau_{nuc}\sim (1\rm GeV)^{-1}$.
In nucleus $\Delta E\sim 100~\rm MeV$\cite{Mohapatra:2009wp}.
Thus in this era eq.(\ref{Pnosci-gen}) becomes approximately:
  $P_{n\rightarrow\bar{n}}\approx\frac{\delta m^2}{\Delta E^2}$.
Thus the washout rate is $\Gamma^{n\rightarrow\bar{n}, \rm late}_W\sim\frac{\delta m^2}{(\Delta E)^2}/\tau_{nuc}$.
Requiring $\Gamma^{n\rightarrow\bar{n}, \rm late}_W<H_0$, we find $\delta m\lesssim 10^{-22}\rm GeV$.
\underline{Current day precision tests:}\\
$n-\bar{n}$ oscillation reactor experiments today set a bound $\delta m\leq 6\times10^{-33}\rm GeV\approx(10^{8}\rm sec)^{-1}$\cite{Mohapatra:2009wp}, 
which is stronger than the washout constraints above. Now we consider constraints from $\delta m$ on model parameters $\lambda_{ij}$. In this minimal model, $\lambda_{ij}$ for $\phi d_i d_j$ have to be anti-symmetric in $i, j$. Consequently $uddudd$ operator giving rise to $\delta m$ is highly suppressed, and $\lambda_{ij}$ are not effectively constrained by $n-\bar{n}$ oscillation\cite{Goity:1994dq}. More relevant constraint comes from $pp\rightarrow K^+K^+$ decay via higher dimensional $\cancel{B}$ operator, which gives bound $\lambda_{12}\lesssim10^{-7}$ for $m_\phi, m_\psi\sim1$ TeV\cite{Goity:1994dq}. As we will show later, when embedding this model in natural SUSY where additional fields such as $\tilde{d}_i$ and related interactions are involved, $n-\bar{n}$ oscillation gives strong bound on $\lambda$-type couplings. We are also constrained by flavor changing neutral currents such as $D_0-\bar{D}_0$ mixing, which gives $y_1y_2\lesssim10^{-2}$ with TeV masses. The large $\epsilon_{CP}$ required for baryogenesis may bring additional constraints from the \textit{neutron EDM}. If $\epsilon_{CP}$ comes from an $O(1)$ phase in $m_\psi, y_i$, in the minimal model where new couplings only involve RH $u_i$, then the contribution involving external quarks vanishes at 1-loop for a similar reason as in the SM\cite{{Maiani:1975in},{Ellis:1976fn}}; the dominant contribution then arises from the Weinberg operator at 2-loop, which still allows phase up to $1/3$ for $O(1)$ couplings and TeV masses\cite{Brust:2011tb}. An even safer option is to have large $\epsilon_{CP}$ come from phases $m_\chi, \varepsilon_i$, so that the EDM is safely suppressed by $\sim\frac{\varepsilon^2}{16\pi^2}\lesssim10^{-18}$. \\
\indent Now we have seen that precision constraints require the new couplings to the first two generations of quarks need to be suppressed. A simple solution is to consider a third-generation dominated pattern where the new fields couple mostly to $b, t$, with CKM-like suppressions to light quarks. This choice further strongly suppresses the earlier washout.
\section{SUSY incarnation and phenomenology}
Now consider an incarnation of our minimal model in ``natural SUSY'' \cite{Barbieri:1987fn} framework with $\cancel{B}$ RPV couplings\cite{Brust:2012uf}.
We promote singlets $\chi$ and $S$ to chiral superfields which we add to the MSSM.
Superpotential terms relevant to our setup are:
\bea
  \nonumber W&\supset& \lambda_{ij}TD_iD_j + \varepsilon' \chi H_uH_d + y_t QH_uT++\mu_\chi\chi^2\\
  &+& \mu H_uH_d+\mu_SS^2 + \alpha\chi^2S+ \beta SH_uH_d.\label{susymodel}
\eea
We assume SUSY breaking such that scalar component of $\chi$ and the first two generation squarks are heavy and decouple from the low energy spectrum,  as in ``natural SUSY''. The diquark $\phi$ in our minimal model is identified with the light $\tilde{t}_R$ in superfield $T$, Majorana $\psi$ is identified as a gaugino (Dirac Higgsino mass is not $\cancel{B}$).
  In eq.(\ref{susymodel}) the terms in the first line ensures $\cancel{B}$ and $\cancel{CP}$ in $\chi$ decay, the $\mu$-terms give masses to fermions and also induce $S-H_u$ mixing which enables a promising channel for LHC search as we will discuss later, and the last two trilinear terms involving $S$ provide WIMP annihilation for $\chi$. $\varepsilon'$ is a reflection of the $\varepsilon$ in our non-SUSY model, enabling late decay $\chi\rightarrow\tilde{\bar{t}} t$ via $\chi-\tilde{H}_u$ mixing. Most of our earlier analysis for the non-SUSY model directly applies here, except for effects from additional fields and interactions. Here gaugino $\psi$ has both LH and RH couplings. Therefore if $\epsilon_{CP}$ is from a gaugino, the 1-loop neutron EDM with external quarks is non-vanishing, but is well suppressed with third generation-dominated flavor pattern\cite{Brust:2011tb}. $n,\bar{n}$ oscillation now constrains $\lambda_{12}, \lambda_{31}\lesssim10^{-3}$, but $\lambda_{23}$ could be $O(1)$\cite{Chemtob:2004xr}, which are again naturally satisfied with third generation dominance. \\
\indent RPV $\cancel{B}$ natural SUSY is intriguing in both theoretical and experimental aspects. However, this scenario suffers from a cosmological crisis. Assuming an otherwise successful conventional baryogenesis at or above EW scale, RPV strong enough for prompt decays within the LHC would typically wash out any primordial B-asymmetry\cite{Barbier:2004ez}. Our SUSY model serves as a robust cure to this problem by having baryogenesis below the weak scale when all wash-out effects decouple. 
To see the problem clearly, as shown in \cite{Brust:2012uf}, for a natural stop that dominantly decays by $\cancel{B}$ couplings, $\lambda_{ij}\gtrsim10^{-7}$ is required to have prompt decay at collider, i.e. decay length $L\lesssim 1\rm mm$. On the other hand, $\lambda_{ij}\gtrsim10^{-7}$ happens to be the range where $\cancel{B}$ scattering such as $\tilde{H}_u t\rightarrow d_id_j$ can efficiently destroy pre-existing B-asymmetry $Y_B^{\rm init}$\cite{Barbier:2004ez}. A simple estimate of such wash-out effect can be read off by dropping the first term at RHS of eq.(\ref{Bsol}). With $\Gamma_{\rm W}\sim\lambda_{ij}^2y_t^2T$, we find an exponential reduction $Y_B(T\approx0)\sim Y_B^{\rm init}e^{-\frac{\lambda_{ij}^2y_t^2}{g_*^{1/2}}\frac{\mpl}{m_{\rm EW}}}$.

\underline{LHC Phenomenology}\\
A promising channel is single resonance production of a mostly-singlet heavy scalar admixture of $H$ and $S$ which dominantly decays to $\chi\chi$. The production channels are the same as for the SM Higgs, except for a mixing suppression. At 14 TeV LHC run, a Higgs-like boson can be produced copiously, even when it is as heavy as 800 GeV, with say $10\%$ mixing, $\sigma\sim 10~\rm fb$.  The produced $\chi$ must live beyond its freezeout, so its lifetime $\tau_{\rm D}\gtrsim t_{\rm f}\sim (1\rm sec)\left(\frac{\rm MeV}{T_{\rm f}}\right)^2>1\rm cm$, where $T_f\lesssim100\rm ~GeV$ so that $m_\chi\lesssim O(\rm TeV)$ is within the LHC reach. Close to this bound on $\tau_D$, $\chi$ decay leaves a displaced vertex  inside detector involving $t,\bar{t}$. The search can be based on dedicated displaced vertex trigger\cite{ATLAS:2012av,{Graham:2012th}}, or triggered on two tagging jets in VBF production channel. A challenging but exciting further step is to measure $\cancel{CP}$ responsible for baryogenesis from the charge asymmetry in the $t \bar{t}$ system.
\section{Summary/Outlook}
We proposed a new mechanism addressing $\Omega_{\rm DM}-\Omega_B$ ``coincidence'' while preserving the merits of the WIMP miracle, presenting a simple example model as well as its incarnation in $\cancel{B}$ natural SUSY. Our basic idea allows for further elaborations, e.g. the WIMP parent may decay to both asymmetric DM and baryons, or baryogenesis may proceed through 3-body decay, accommodating a lighter $\chi$.  In mini-split SUSY \cite{Wells:2004di, Arvanitaki:2012ps, Feldstein:2012bu}  the latter has a natural incarnation\cite{cui}. On the phenomenology side, our mechanism  brings the exciting possibility of having the cosmological origin of matter being testable at current-day colliders. It is also possible that with improvements low energy experiments will be another frontier to test the mechanism we proposed.

\section*{Acknowledgments}
We thank C.~Brust, D.~E.~Kaplan, A.~Katz, R.~Mohapatra, J.~Ruderman, B.~Shuve and L.~Vecchi for helpful discussions.  Y.~Cui and R.~Sundrum are supported in part by NSF grant PHY-0968854 and by the Maryland Center for Fundamental Physics. R.~Sundrum was also supported by NSF grant PHY-0910467 .

\end{document}